\documentclass[useAMS,usenatbib]{mn2e}
\usepackage{natbib}
\usepackage{graphicx}
\usepackage{amsmath}
\usepackage{amssymb}
\usepackage{color}
\usepackage{hyperref}

\DeclareMathOperator{\sign}{sign}
\DeclareMathOperator{\abs}{abs}

\newcommand{\uA}{\mathrm{A}}
\newcommand{\uM}{\mathrm{M}}
\newcommand{\uS}{\mathrm{S}}

\newcommand{\ug}{\mathrm{g}}

\newcommand{\apj}{ApJ}
\newcommand{\aap}{A\&A}
\newcommand{\apjl}{ApJL}
\newcommand{\mnras}{MNRAS}

\newcommand{\nat}{Nature}

\title[Relativistic MHD jets with gravity]{Linking accretion flow and particle acceleration in jets. I. New relativistic magnetohydrodynamical jet solutions including gravity}

\author[Peter Polko, David L. Meier and Sera Markoff]{Peter Polko$^{1}$\thanks{E-mail: P.Polko@uva.nl}, David L. Meier$^{2}$ and Sera Markoff$^{1}$\\
$^{1}$Astronomical Institute `Anton Pannekoek', University of Amsterdam, P.O. Box 94249, 1090 GE Amsterdam, the Netherlands\\
$^{2}$Jet Propulsion Laboratory, California Institute of Technology, Pasadena, CA 91109, USA}
\begin{document}

\pagerange{\pageref{firstpage}--\pageref{lastpage}} \pubyear{2012}
\maketitle

\begin{abstract}
We present a new, approximate method for modelling the acceleration and collimation of relativistic jets in the presence of gravity. This method is self-similar throughout the computational domain where gravitational effects are negligible and, where significant, self-similar within a flux tube. These solutions are applicable to jets launched from a small region (e.g., near the inner edge of an accretion disk). As implied by earlier work, the flow can converge onto the rotation axis, potentially creating a collimation shock.

In this first version of the method, we derive the gravitational contribution to the relativistic equations by analogy with non-relativistic flow.

This approach captures the relativistic kinetic gravitational mass of the flowing plasma, but not that due to internal thermal and magnetic energies. A more sophisticated treatment, derived from the basic general relativistic magnetohydrodynamical equations, is currently being developed.

Here we present an initial exploration of parameter space, describing the effects the model parameters have on flow solutions and the location of the collimation shock. These results provide the groundwork for new, semi-analytic models of relativistic jets which can constrain conditions near the black hole by fitting the jet break seen increasingly in X-ray binaries.
\end{abstract}


\section{Introduction}
\label{firstpage}
Despite decades of study, astrophysical jets are still an enigmatic component of our universe. Understanding the details of their physics would advance fields as diverse as star formation, gamma-ray bursts (GRBs), galaxy evolution and cluster dynamics. Jets in all of these objects probably share three basic ingredients: a source of energy and particles such as an accretion flow, magnetic fields created or carried inward by the flow, and rotation. However, the exact details of how these elements combine to launch and collimate jets in the observed systems is not at all clear.

Black holes are ideal test cases to understand jet physics because of their enormous range in mass, allowing us to study jet dynamics over an equally large range of time-scales. Accreting stellar mass black holes in a binary with a companion star (BHBs) have jets that can undergo an entire launch and quench cycle on time-scales of months, which is related at least in part to the accretion state of the disc \citep{1997ApJ...489..865E,2001ApJ...548L...9M,2004MNRAS.355.1105F,2006csxs.book..157M}. In their supermassive cousins, active galactic nuclei (AGN), such transitions would be expected to occur on time-scales of millions to a billion times longer. To assess the extent to which such a mass-scaling operates, we must have a better understanding of which elements of jet physics persist despite a dramatic range of environment and scales.

One key feature of jets regardless of source seems to be their ability to accelerate particles to highly relativistic energy. We observe the outcome of these accelerated particles directly via optically thin synchrotron emission in, e.g., AGN \citep{1985ApJ...298..114M} and in BHBs \citep{2001MNRAS.322...31F}. When the jets are compact enough to become self-absorbed, the superposition of many synchrotron-emitting components along the length of the jet leads to a flat or slightly inverted total spectrum \citep{1979ApJ...232...34B} until the point where the jet emission becomes entirely optically thin. At this point the spectral index will steepen dramatically to $\alpha \sim 0.5-0.8$ typically, where $\alpha$ is defined such that $F_\nu \propto \nu^{-\alpha}$. This break in the spectrum can be physically associated with the most compact region in the jets where particle acceleration is present. In low-luminosity AGN jets this break typically occurs in the GHz range \citep{1999ApJ...516..672H}, and if jets are self-similar across the black hole mass range, in BHBs such a break would be predicted to occur in the infrared (IR) bands \citep{2001A&A...372L..25M,2003A&A...397..645M,2003MNRAS.343L..59H}. Because the spectral break also indicates the total radiative power of the jets, and plays a key role in driving mass-scaling predictions such as the Fundamental Plane of black hole accretion \citep{2003MNRAS.345.1057M,2004A&A...414..895F,2012MNRAS.419..267P}, it is a pivotal parameter linking the inner jets (and assumedly conditions near their launch point) to the outer jets associated with strong radio through IR emission. Deriving this link and overall jets structure in an MHD-consistent way is the focus of this paper.

Constraints on the location of the jet break can come from both observations as well as from spectral fitting. Because the optical/IR (OIR) bands are often dominated by a stellar companion or the accretion disc, the optically thick-to-thin break has been observed explicitly in only one BHB so far, GX~339-4 (\citealt{2002ApJ...573L..35C, 2011ApJ...740L..13G}; see also \citealt{2006ApJ...643L..41M} for a similar break for neutron stars), and has been indirectly constrained for Cyg~X-1 based on recent MIR observations \citep{2011ApJ...736...63R}. These observations confirm the theoretically predicted break location, and the recent boom in multiwavelength monitoring of many BHBs is leading to new studies of the break and its scaling with luminosity (Russell et al., in prep.). The location in frequency of the break can also be rather tightly constrained by fitting multiple spectra of BHBs and AGN in states associated with compact jets, which has the advantage of also associating a size scale for the region, and distance from the black hole. A series of works have consistently found the region of the jet where particle acceleration initiates seems to be offset from the black hole at distances of $\sim$10--1000$~r_\ug$, depending on source luminosity, where $r_\ug$ is the gravitational radius, $GM/c^2$ \citep{2001A&A...372L..25M,2003A&A...397..645M,2005ApJ...635.1203M,2008ApJ...681..905M,2007ApJ...670..610M,2007ApJ...670..600G,2009MNRAS.398.1638M}. Accordingly, this zone also should be associated with an offset in the start of the synchrotron core from the black hole in direct imaging. Such a measurement is very difficult to do because of the spatial resolution required. At least for the jet in M87, a relatively nearby AGN \citep[$17.0 \pm 0.3$~Mpc,][]{2001ApJ...546..681T} with a large supermassive black hole \citep[$6.4 \pm 0.5 \times 10^{9}~$M$_\odot$,][although note this value is twice the mass found in previous studies]{2009ApJ...700.1690G}, the offset is $\sim 30 \textrm{--} 100~r_\ug$ \citep{1999Natur.401..891J,2008JPhCS.131a2053W,2011Natur.477..185H}. Interestingly, recent astrometry for M87 has also shown that the location of the radio core with respect to the black hole is extremely stable over several years \citep{2011IAUS..275..198A}. A natural explanation for such a region for the start of particle acceleration would be a shock where continual diffusive acceleration occurs \citep[e.g.][]{1978MNRAS.182..147B,1983RPPh...46..973D}. Because particles need to be continually accelerated, the shock should be a steady feature within the compact jets.

In self-similar, axisymmetric MHD flows, the bulk flows can be accelerated through several singular points in order: the modified slow point (MSP), the Alfv\`en point (AP), and the modified fast point (MFP) \citep[hereafter BP82]{1982MNRAS.199..883B}. At the MFP, the velocity of the flow greatly exceeds the fast magnetosonic speed, which is the fastest that signals in the jet can travel. Thus beyond the MFP the flow is causally disconnected from the flow closer to the central object, allowing disruptions such as shocks to form. A stable location for the MFP in a given flow would therefore provide a natural explanation for a steady feature associated with particle acceleration. This location could, however, be a function of mass and accretion power and vary from source to source within a certain range. In the models mentioned above, the shock location is a fitted free parameter that suggestively falls into the same range for a variety of sources. However, to understand if this is a physically meaningful result, we should ideally be able to show that this location corresponds to a feature that can be derived self-consistently within a theoretical framework, and tied to conditions at the jet launch point. The earlier modes are based on a hydrodynamical (HD) velocity profile \citep{1995A&A...293..665F,2000A&A...362..113F}, and thus do not invoke the role of magnetic fields other than as a global parameter. However, if we move to a MHD framework, we can derive the exact location of the MFP (and assumedly the shock) based on boundary conditions defined by the launch point of the jets, and then test if the theory provides the correct scalings as observed, and as seem to be driving the Fundamental Plane.

In this paper we take a semi-analytical approach to derive new flow solutions (a new jet model) with the explicit aim of exploring the conditions for formation of the MFP, by solving the equations of relativistic MHD under the assumptions of self-similarity and axisymmetry. These assumptions greatly simplify the equations describing the jets, and are consistent with the results of numerical MHD simulations, which almost always produce a magnetic field line geometry near the launch point that is remarkably self-similar and axisymmetric \citep[see, e.g., fig.~2 and fig.~11 in][]{2006MNRAS.368.1561M}.

We need to make one further approximation to construct our solutions: in order to link the properties of the MFP to the conditions at the base of the jet, we need to solve for the AP and MSP in a regime which could potentially be very close to the black hole where gravity becomes important. The MSP is a natural point to associate with the launch region of the jets, similar to the sonic nozzle in hydrodynamical flows, and is thus also the place where conditions could eventually be matched to the accretion inflow, such as a magnetic corona or radiatively-inefficient accretion flow \citep[RIAF;][]{1994ApJ...428L..13N,1999MNRAS.303L...1B,2000ApJ...539..809Q,2002MNRAS.332..165M}. However, gravity will play a significant role so close to the black hole; therefore, it also needs to be included in order to correctly derive and cross the MSP. \emph{Until now no semi-analytical formalism has been developed to describe a relativistic flow passing through all three singular points, because in a relativistic framework gravity is not self-similar}. Therefore in this paper, in order to accomplish this connection, we apply a bridging solution between our previous relativistic self-similar flow model without gravity (\citealt[][hereafter PMM10]{2010ApJ...723.1343P}, based on the framework for self-similar flow laid out in \citealt[][hereafter VK03]{2003ApJ...596.1080V}), which is also valid in the non-relativistic limit (again without gravity), and a non-relativistic flow model that is valid close to the black hole and that does include the effects of gravity self-consistently \citep[][hereafter VTST00]{2000MNRAS.318..417V}. The transition from the non-relativistic flow with gravity to our relativistic flow without gravity from PMM10 occurs sufficiently far from the black hole that gravitational effects are no longer important. The bridging of these two solutions will not describe all possible gravitational fields and velocities, but the range of solutions is wide enough to be applicable to many observed sources, which is the goal of this project. Specifically, within a cylindrical region constituting a ``flux tube'', that can be associated with the observable ``sheath'' of compact jets, we demonstrate that the deviations from self-similarity when gravity is present are very small. Within this scenario, we present relativistic MHD solutions that cross all three singular points, allowing us to derive the location of the shock self-consistently from boundary conditions at the base of the jet and to begin to test how well such a model fares against physical sources.

Gravity is treated in the weak (Newtonian) limit, and in this paper we have opted for a Paczy{\'n}sky-Wiita (pseudo-Newtonian) augmentation since some of our solutions approach the Schwarzschild radius, where general relativistic effects begin to be important. No other GR correction terms, nor black hole spin, is treated at this time.

In \S 2 we present details of the bridging formalism. In \S 3 we use the new formalism to obtain solutions crossing the MSP, AP, and MFP and describe their properties. We also show the effects that changing the model parameters have on the location of the MSP and MFP. In \S 4 we present an in-depth discussion of our model and results, and we draw our conclusions in \S 5.

\section{Method}
In order to tie the conditions at the start of the particle acceleration region to the conditions at the base of the jet, we need to have a full description of a hot, relativistic flow. Because close to the central object gravity is as important as temperature and the magnetic field strength, the gravitational potential must be taken into account in order to be able to describe the jet in this region. Because this region is generally subrelativistic or mildly relativistic, we start with a solution that includes gravity in a non-relativistic flow. Far from the black hole the flow will be relativistic, but gravity will no longer be important in the solution, so we can use the equations derived in PMM10 to describe this flow. We then seek solutions that satisfy both the relativistic flow without gravity and the non-relativistic flow with gravity at opposite ends of the jet.

\subsection{A physical description of the flow}
\label{physicalflow}
In the framework of self-similar relativistic MHD, we find that jets in our solutions are accelerated in the following way: starting from the base of the flow, the initial bulk acceleration is provided through thermal pressure gradients by means of sound waves by the very hot particles surrounding the central object. When the flow exceeds the sound speed, this mechanism becomes inefficient, and acceleration due to the centrifugal force of the rotating magnetic field by means of Alfv\'en waves takes over. After the flow surpasses the Alfv\'en speed, magnetic pressure becomes the dominant mechanism. When the flow velocity increases beyond the fast magnetosonic speed, the final boost is given by the pinching of the magnetic field. If at this point the flow overcollimates, a shock may form, which can accelerate the particles into the observed power-law distributions, thus constituting the start of the particle acceleration region.

In prior work (PMM10) we were unable to probe the region where the initial bulk acceleration occurs due to the absence of gravity in our formalism. However, the singular MSP can be produced only by the inward gravitational force balancing the outward magnetic and thermal pressure forces. Thus by including gravity we can describe the jet from very close to the central object to the first instance of overcollimation, providing the connection between the conditions at these two points. We will discuss how we deal with the issues regarding self-similarity in \S \ref{discussion}.

\subsection{A new solution technique: the \texorpdfstring{$C^\infty$}{C infinity}-continuous bridging method}
\label{bridgingsolution}
The matching of two flow solutions valid in different regimes of parameter or physical space is a common technique in physics. Some techniques simply solve the two different equations in the different regimes and then match the solutions at a single (and arbitrary) point. Such bridging solutions are only $C^0$-continuous, meaning its derivative is not a continuous function. Other methods may involve spline fitting the two solutions together in a finite \emph{bridging region}. In this case, the two solutions are valid in each of their respective regimes, but in the bridging region neither is strictly valid. Instead, the spline fit is an approximation of the transition between the two solutions, and it is $C^1$-continuous or better.

Our method for bridging the non-relativistic VTST00 MHD wind equation with gravity and the relativistic VK03 one without gravity is similar to the above approaches. That is, when gravitational forces are important and the flow non-relativistic, we obtain the VTST00 MHD solution. However, when gravity is no longer important, the flow has the character of the VK03 relativistic solutions, whether it is relativistic or not. The difference here is that, because of the nearly-identical structures of the VTST00 and VK03 equations, we can perform this task with the construction of a \emph{single} wind equation. The method, then, will be $C^\infty$-continuous, with all derivatives being continuous functions -- a distinct advantage when one is dealing with equations which have singular points through which the solution must pass.

\subsection{The basic \texorpdfstring{$C^\infty$}{C infinity}-continuous method}
\label{gravityderivation}
We now give a brief overview of our previous work in PMM10 and describe how we here extend it by bridging it with a non-relativistic formulation including gravity valid close to the black hole.

In PMM10 we combined the Bernoulli equation (describing the forces along a field line) and the transfield equation (describing the forces perpendicular to a field line), to construct a wind equation, which describes the bulk acceleration of the flow and fully specifies the jet geometry. Using this single differential equation, it was possible to obtain jet solutions that crossed an MFP at a finite distance from the origin. The terms in the wind equation thus obtained using the formalism of VK03 turned out to be almost identical to the terms in the wind equation given in VTST00. Apart from relativistic corrections to all the corresponding terms, the only difference is the gravity term in the VTST00 wind equation, which is non-relativistic. If we modify the VK03 wind equation to include gravity, our combined wind equation will reduce exactly to the formalism of VTST00 when relativistic effects apart from gravity are negligible as is the case close to the black hole, while it also reduces exactly to the formalism of VK03 when gravity becomes negligible as is the case far away from the black hole. Between these two regimes the wind equation describes the bridging regime where relativistic effects may begin to dominate the gravitational effects. Since all these regimes are described by one continuous equation, the resulting solution will be a smooth transition between the two regimes. We prefer this approach over matching the solutions of the two regimes at an arbitrary location in the jet, while trying to satisfy any number of continuity conditions.

Since the notation in VTST00 differs from VK03, we use equations (8)--(12) in VTST00 and equations (24) in VK03 to translate the notation to that used in the latter. This conversion also provides the relativistic corrections to the gravitational mass. Using this relativistic velocity in the definition of the gravity term in VTST00 yields a relativistic form of the gravity term, allowing us to include it into the VK03 wind equation. The term by term comparison of the two wind equations ensures that we use the proper scaling. Since some of our solutions get very close to the central black hole, we adapted the gravity term to include a pseudo-Newtonian potential \citep{1980A&A....88...23P}.

After these steps (see appendix \ref{appendix:gravity}) the gravity term has the following form:
\begin{equation}
- \frac{\mu^2 x_\uA^4}{F^2 \sigma_\uM^2} \frac{(1 - M^2 - x_\uA^2)^2}{(1 - M^2 - x^2)^2} \left[ \frac{c^2 \varpi_\uA G}{\mathcal{G M} \sin(\theta)} - 2 \right]^{-1},
\label{gravityterm}
\end{equation}
where the subscript A denotes values at the Alfv\'en point and $\mu c^2$ is the total energy-to-mass flux ratio, $x_\uA$ is the cylindrical radius distance of the AP scaled to the light cylinder radius, $F$ controls the current distribution, $\sigma_\uM$ is the magnetisation parameter, $M$ is the Alfv\'enic Mach number, $x$ is the cylindrical radius distance scaled to the light cylinder radius, $c$ is the speed of light, $\varpi_\uA$ is the cylindrical radius distance to the AP, $G$ is the cylindrical radius distance scaled to the AP, $\mathcal{G}$ is the gravitational constant, $\mathcal{M}$ is the mass of the black hole and $\theta$ is the spherical polar angle (see Table \ref{parameters} for an overview of all model parameters).

The first two fractions consist purely of constants along a given field line, providing the overall scaling. The third fraction is proportional to $(\gamma \xi)^2$ (see \ref{kappa}), where $\gamma$ is the Lorentz factor and $\xi c^2$ is the specific relativistic enthalpy (per baryon mass). The last fraction corresponds to $(r/r_\ug - 2)^{-1}$, with $r$ the spherical radius and $r_\ug$ the gravitational radius, and it is this term that represents the pseudo-Newtonian potential.

Several equations besides the wind equation are needed to fully determine a solution, so we have made sure that we take into account all equations in which the gravity term appears. As explained above, gravity, as described by equation \eqref{gravityterm}, shows up in the numerator of the wind equation. By evaluating the wind equation at the AP, we obtain the Alfv\'en regularity condition (ARC, see equation \ref{ARC}), which also depends on gravity. The ARC allows us to calculate the slope of $M^2$ at the AP, $p_\uA$, one of the initial parameters of the integration. The gravity term, as it appears on the right hand side of the ARC in VK03, is given by:
\begin{equation}
- \frac{x_\uA^2}{1 - x_\uA^2} \left[ \frac{c^2 \varpi_\uA}{\mathcal{G M} \sin(\theta_\uA)} - 2 \right]^{-1}.
\end{equation}

Another location in which the gravity term may appear is in the Bernoulli equation evaluated at the AP. This equation is used to determine $\mu$, another initial parameter of the integration (see equation \ref{mu2}). However, by evaluating the Bernoulli equation of VTST00 at the AP, the gravity term in that equation vanishes there. So equation (B5) in VK03 is still valid (see appendix \ref{Bernoulli}). The transfield equation is incorporated into the wind equation, but not used independently in our calculations, so its dependence on gravity is irrelevant here.

By making these changes we have been able to find solutions that can pass through all three singular points of a hot, relativistic MHD flow, extending all models so far published (for a qualitative overview see Table \ref{modeloverview}).

\begin{table*}
\caption{Overview of self-similar, axisymmetric MHD models, classified whether they include gravity, allow for a warm flow, allow relativistic velocities, and cross the MSP, AP and MFP, respectively.}
\begin{center}
\begin{tabular}{|l||c|c|c||c|c|c|}
\hline
\hspace{\stretch{1}} Paper \hspace{\stretch{1}} & Gravity & Warm & Relativistic & MSP & AP & MFP \\
\hline
\citet{1982MNRAS.199..883B} & \checkmark & & & & \checkmark & \\
\citet{1992ApJ...394..459L} & & & \checkmark & & \checkmark & \\
\citet{2000MNRAS.318..417V} & \checkmark & \checkmark & & \checkmark & \checkmark & \checkmark \\
\citet{2003ApJ...596.1080V} & & \checkmark & \checkmark & & \checkmark & \\
\citet{2010ApJ...723.1343P} & & \checkmark & \checkmark & & \checkmark & \checkmark \\
Polko et al.~(this work) & \checkmark & \checkmark & \checkmark & \checkmark & \checkmark & \checkmark \\
\hline
\end{tabular}
\end{center}
\label{modeloverview}
\end{table*}

\subsection{The effects of including gravity on the solutions}
\label{gravityeffects}
In every instance of gravity the ratio $\varpi_\uA / \mathcal{M}$ appears. Thus for a central object that is twice as massive, if the AP is moved twice as far from the axis of symmetry, the solution remains unchanged. This result is a direct consequence of mass-scaling, since all properties are expressed in gravitational radii. Regulating the gravitational force exerted by the compact object can therefore be achieved by changing either $\mathcal{M}$ or $\varpi_\uA$ and, apart from the overall physical size of the system, these two actions are equivalent. Increasing the effect of gravity can thus be thought of either as increasing the mass of the central object (increasing the reach of the gravitational well), or as decreasing the radius of the AP (moving the Alfv\'en point deeper into the gravitational well).

One way to show these effects, would be by starting with a singular solution containing both an MFP and an MSP, slowly increasing $\varpi_\uA$ to decrease the effect of gravity, while keeping solutions with an MFP and MSP. Unfortunately, these solutions do not smoothly transition into solutions that only have an MFP. It is therefore impossible to directly compare these new solutions crossing all three singular points to solutions without an MSP and we are left noting the effects within these new solutions.

The significance of having a solution with an MSP is that only then it is possible to connect conditions at the MFP to conditions very close to the central object. Even though the extension of a field line in linear distance may not be much, we can be sure that the results in the sub-Alfv\'enic regime have meaning by satisfying a strict boundary condition. We equate the MSP with the launch point of the jet and fit the conditions at the MSP to the accretion flow. Hence, extending the solutions to the MSP allows a link between the accretion flow and the start of the particle acceleration region, which we will use in later work.

Compared with the Paczy{\'n}sky-Wiita potential, the Newtonian potential has a weaker gravitational force at the same radius. As it is the gravitational force that balances all other forces at the MSP, using a Newtonian potential would move the MSP closer to the black hole, possibly beyond the event horizon for certain solutions. To avoid this, we adopt the Paczy{\'n}sky-Wiita potential to approximate the gravitational potential close to the black hole. The adaptation from a Newtonian potential is straightforward (see appendix \ref{appendix:gravity}).

\subsection{Model parameters}
\begin{table}
\caption{List of model parameters}
\begin{center}
\begin{tabular}{@{}ll@{}}
\hline
\multicolumn{2}{c}{Fundamental parameters}\\
\hline
$F$ & Sets the radial dependence of the magnetic field strength \\
$\Gamma$ & Adiabatic index \\
$x_\uA$ & Cylindrical radius in terms of the light cylinder radius \\
$\sigma_\uM$ & Magnetisation parameter \\
$q$ & Dimensionless adiabatic coefficient \\
$\varpi_\uA$ & Cylindrical radius \\
$\mathcal{M}$ & Mass of the central object \\
\hline
\multicolumn{2}{c}{Derived parameters}\\
\hline
$\theta_\uA$ & Angle with respect to the axis of symmetry \\
$\psi_\uA$ & Angle of the field line with respect to the disc \\
$p_\uA$ & Derivative of the Alfv\'enic Mach number squared w.r.t. $\theta$ \\
$\sigma_\uA$ & Magnetisation func. (Poynting-to-matter energy flux ratio) \\
$\mu c^2$ & Total energy-to-mass flux ratio \\
$\xi c^2$ & Specific (per baryon mass) relativistic enthalpy \\
\hline
\end{tabular}
\end{center}
\label{parameters}
Notes: The subscript A means the value of the variable at the AP. For a complete description of the parameters, please see PMM10.
\end{table}

By including gravity in the equations, two additional parameters need to be specified before an integration. These are the mass of the central object, $\mathcal{M}$, and the cylindrical radius of the AP, $\varpi_\uA$. As mentioned above, only the ratio of these parameters appears in the equations, so these two parameters are linearly dependent and solutions with the same ratio are indistinguishable if all other parameters are fixed. Since a solution is given in gravitational radii, this means that any solution can be scaled up or down to any physical size. For our initial solution we have chosen to set $\mathcal{M}$ to the mass of a typical stellar mass black hole of $10~M_\odot$ (although this value will eventually be set to the deduced mass of an observed black hole system) and to set $\varpi_\uA$ to $5~r_\ug$. We will look at the effects of mass scaling at a later time.

The parameters $F$ and $\Gamma$ are set to a single value for now, but can be changed if necessary. For example, since $F$ influences the geometry of the jet, it also affects the collimation. By allowing it to vary, we will have the freedom to adjust the degree of collimation to fit the broad-band data on our sources. Due to the self-similar nature of the equations, the height of the MFP will depend on the initial radius at the disc. Because we want a representative location, we choose to anchor the field line in the most active part of the accretion disc close to the central object. Within the innermost stable circular orbit (ISCO) the accreting matter can no longer form a disc and will move towards the central object as a radiatively inefficient flow. Therefore we set $F = 0.75$, which corresponds to a wide range of accretion disc models, including the Blandford-Payne (BP82), Shakura-Sunyaev \citep{1973A&A....24..337S}, and radiatively-inefficient accretion flow (RIAF) models. We also set $\Gamma = 5/3$, appropriate for non-radiation-dominated jets in the hard state, where the radiation and the particles do not behave as a single fluid.

The choice of parameters to fit the singular points will affect the solution we find. Were we to use $\theta_\uA$ and $\psi_\uA$ to fit for the MFP and MSP positions, the angle of the Alfv\'en point and the slope of the field line there would be varied to find a singular solution. If, on the other hand, $x_\uA^2$ and $q$ were used to fit for the MFP and MSP, the light cylinder radius (and hence angular velocity) and dimensionless adiabatic coefficient would be varied instead. This last combination most closely approaches self-similarity out of all the pairs of parameters that we have studied. Therefore, in this work we will use $x_\uA^2$ and $q$ as fitting parameters. It is important to note that by choosing different fitting parameters only the way we move through parameter space changes, not the set of solutions themselves.

\section{Results}
\subsection{First solution}
\label{firstsolution}

\begin{table*}
\caption{Parameters of solutions}
\begin{center}
\begin{tabular}{l c c c c c c c c c c c c}
\hline
& & & $\theta_\uA$ & $\psi_\uA$ & & $\varpi_\uA$ & $\mathcal{M}$ & & & & & \\
& $F$ & $\Gamma$ & (deg) & (deg) & $\sigma_\uM$ & ($r_\ug$) & $M_\odot$ & $x_\uA^2$ & $q$ & $p_\uA$ & $\sigma_\uA$ & $\mu$ \\
\hline
PMM10$^1$ & 0.75 & 5/3 & 50 & 55 & 2.53981 & $-$ & $-$ & 0.75 & 0.12 & $-1.66651$ & 2.60390 & 9.85117 \\
This work & 0.75 & 5/3 & 50 & 60 & 2.5 & 5 & 10 & 0.755500 & 0.271221 & $-1.27754$ & 2.25711 & 12.4036 \\
\hline
\end{tabular}\\
\end{center}
$^1$ The parameters given are for solution $c$ in PMM10. This is the solution we will compare our new result with. The values for the first seven parameters ($F$ through $\mathcal{M}$) of the solution in this work are exact, for the last five ($x_\uA^2$ through $\mu$) they are rounded off. Because singular solutions require high precision, the rounded-off numbers are given with six significant digits.\\
\label{parametertable}
\end{table*}

\begin{figure}
\includegraphics[width = 0.45 \textwidth]{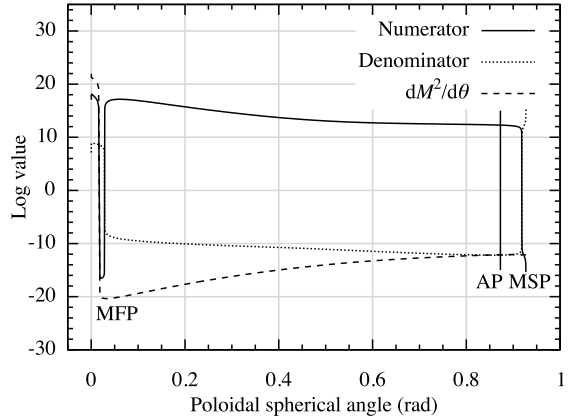}
\caption{
The reference solution showing both an MFP and MSP. The dotted/red and dashed/blue lines show the numerator and denominator of the wind equation respectively, while the solid/black line shows their ratio. The latter determines the total bulk acceleration of $M^2$ with respect to polar angle $\theta$. The vertical line shows the location of the AP. The vertical axis is a `scaled logarithm' of the plotted parameters, i.e., $y = \sign(x) \log_{10} [1 + \abs(x)/10^{-12}]$ to clearly show the variables over many orders of magnitude. At low $\theta$ a small change in angle corresponds to a vast change in height, also contributing to the near vertical change of sign of the bulk acceleration. (A colour version of this figure is available in the online journal.)
}
\label{numden}
\end{figure}

Starting with the warm (initial $\xi \approx 4.7$) and fast (final $\gamma \approx 10$) solution $c$ in PMM10, we use our new method to explore parameter space to determine a solution with an MSP. The parameter values of this new solution are very close to those of solution $c$ (see Table \ref{parametertable}), establishing the relative ease with which solutions can be found. Fig. \ref{numden} shows the numerator and denominator of the wind equation, and their ratio corresponding to the derivative of square of the Alfv\'enic Mach number with respect to the poloidal spherical angle $\theta$. From this ratio it is clear we have indeed found a smooth crossing. In comparison with solution $c$, the AP is still located at $\theta = 0.87$ rad or $50^{\circ}$ from the axis of symmetry, but the MFP has moved outward to $\theta = 0.029$ rad or $1.65^{\circ}$, while the MSP (only ostensible in solution $c$) has moved inward to $\theta = 0.92$ rad or $52.6^{\circ}$. This angle corresponds to a spherical radius of $5.96~r_\ug$, which is just within the ISCO.

\begin{figure*}
\includegraphics[width = 0.45 \textwidth]{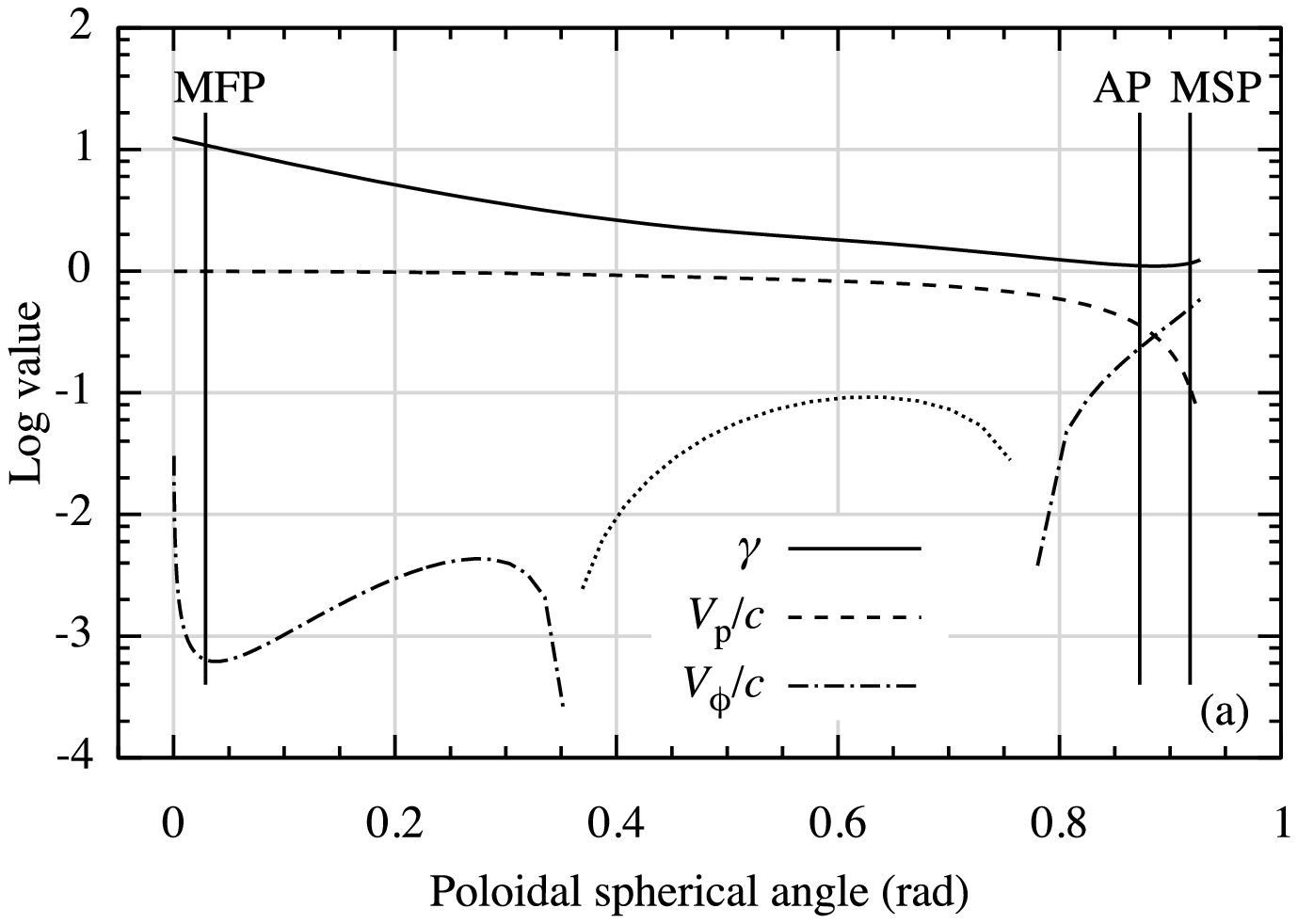}
\includegraphics[width = 0.45 \textwidth]{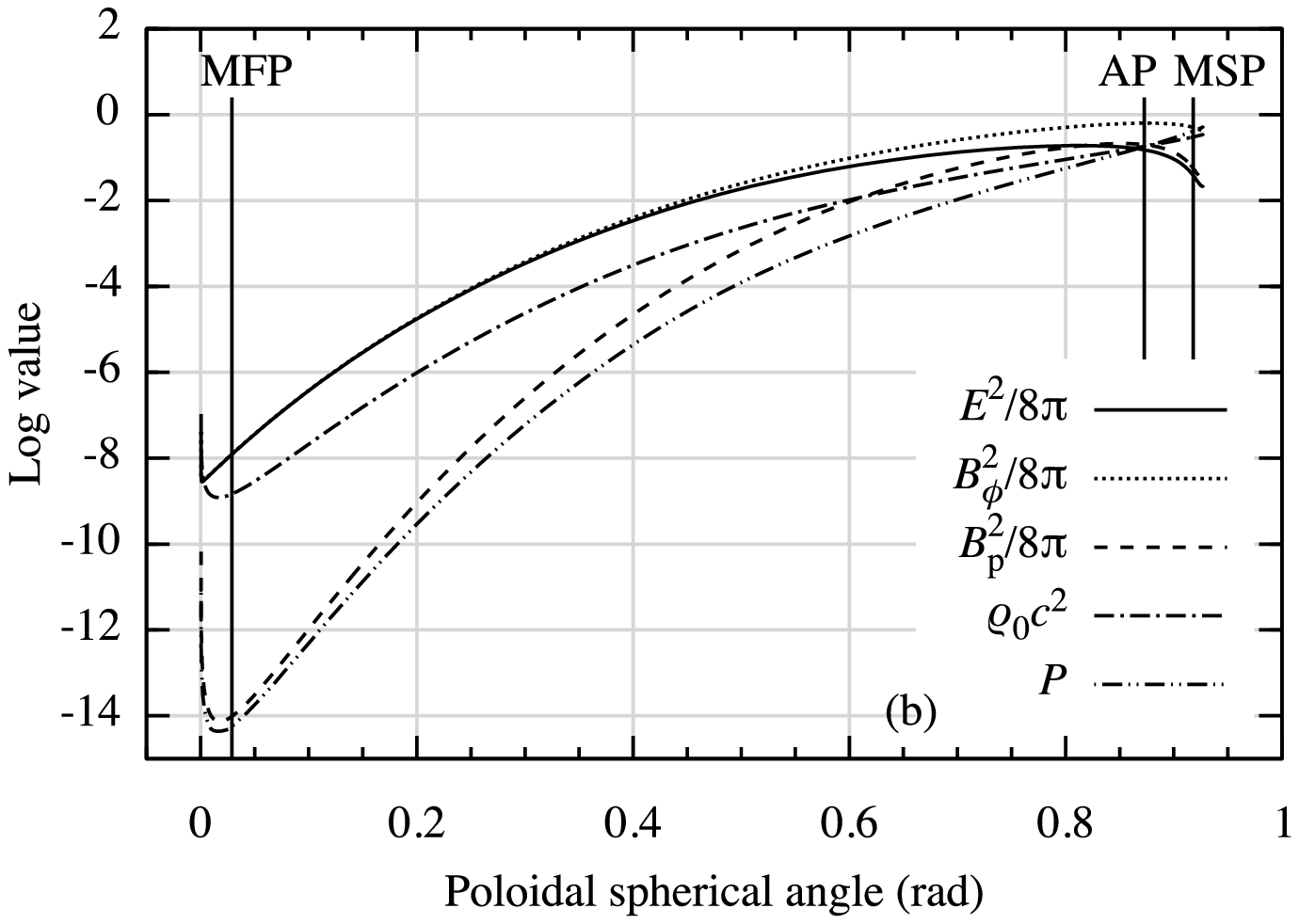}
\caption{
Various physical quantities that result from integrating the wind equation depicted in Fig. \ref{numden}, plotted against polar angle $\theta$. The MFP, AP and MSP are marked. Panel (a) shows the Lorentz factor ($\gamma$), the poloidal ($V_\mathrm{p}$) and toroidal velocity ($V_\phi$) of our canonical solution. Please note that the toroidal velocity starts out positive, turns negative just beyond the AP (the dotted line) and then becomes positive again later in the outflow. Panel (b) shows the electric field energy ($E^2/8\pi$), the toroidal magnetic field energy ($B_\phi^2/8\pi$), the poloidal magnetic field energy ($B_\mathrm{p}^2/8\pi$), the density ($\rho_0 c^2$) and the pressure ($P$). These have all been divided by the scaling factor $B_0^2 \alpha^{F - 2}$, where $B_0$ is a reference magnetic field strength and $\alpha$ is the square of the cylindrical distance of the AP in terms of a reference length ($\alpha = \varpi_\uA^2/\varpi_0^2$), as defined in VK03. Although the square is plotted, $B_{\phi}$ is negative everywhere in this model. (A colour version of this figure is available in the online journal.)
}
\label{physical}
\end{figure*}

The values of the velocity, the Lorentz factor, the magnetic and electric field strength, the density and the pressure along a reference field line are given in Fig. \ref{physical}. The poloidal velocity monotonically increases from $0.067~c$, through $0.11~c$ at the MSP, to very close to the speed of light at the MFP. The toroidal velocity starts off positive, turns negative after the AP, and returns to positive before the MFP. The Lorentz factor slowly increases from 1.17 at the MSP to a final value of 12.3.

The poloidal and (negative) toroidal magnetic field strengths first increase in magnitude and then decrease, both peaking around the AP. The electric field has the same behaviour. The density and pressure both drop monotonically as is expected for an expanding jet. Just beyond the MFP the jet overcollimates, causing both density and pressure to increase again.

\begin{figure*}
\includegraphics[width = 0.95 \textwidth]{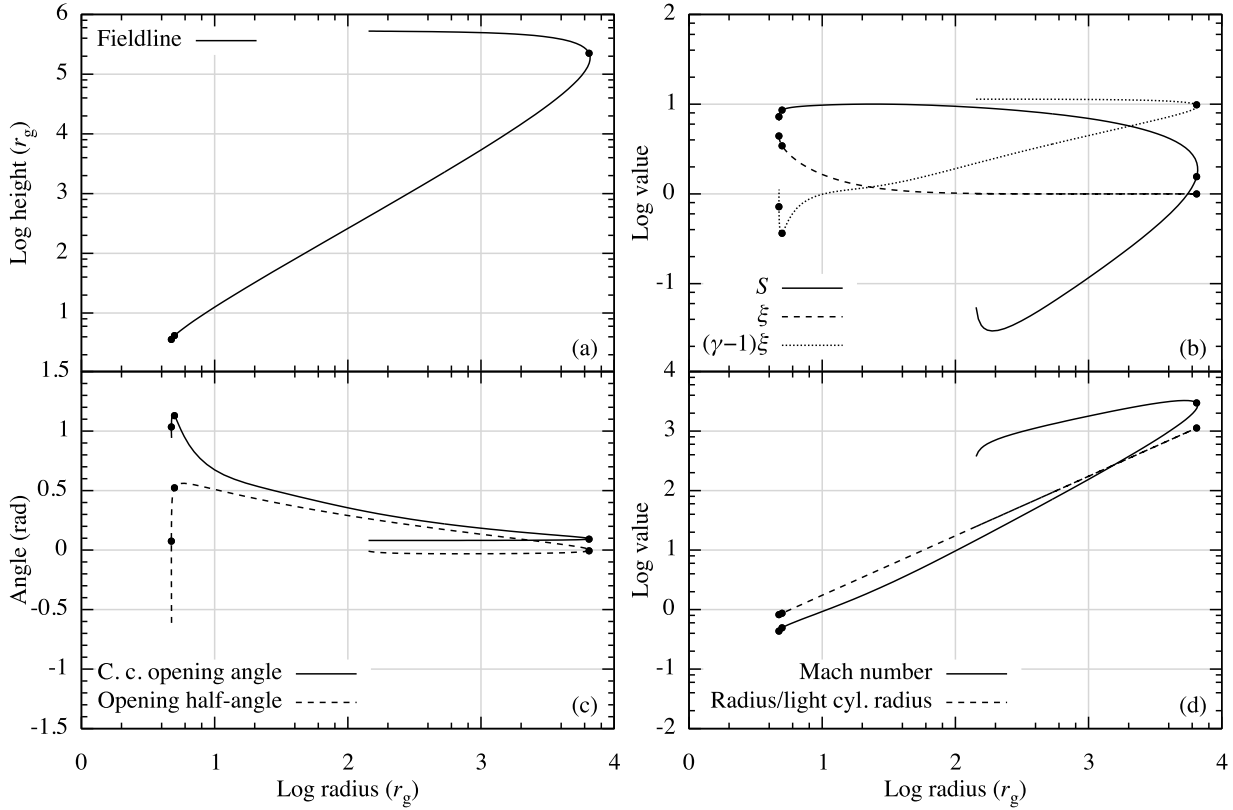}
\caption{
Various physical quantities that result from integrating the wind equation depicted in Fig. \ref{numden}, now plotted against cylindrical radius $\varpi$ instead of $\theta$. These plots are similar to those in fig. 4 of PMM10. Panel (a) shows the geometry of the field line. Panel (b) shows the magnetic energy ($S \equiv - \varpi \Omega B_{\phi} / \Psi_A c^2$), the thermal energy including the rest mass ($\xi$), and the kinetic energy ($[\gamma - 1] \xi$). Panel (c) shows the opening half-angle of the outflow ($\pi / 2 - \psi$) and the causal connection opening angle ($\arcsin [1 / \gamma]$). Panel (d) shows the cylindrical radius in units of the `light cylinder' radius ($x$) and the Alfv\'enic Mach number ($M$). The MSP, AP and MFP (from left to right) are indicated for each quantity. (A colour version of this figure is available in the online journal.)
}
\label{overview}
\end{figure*}

Fig. \ref{overview} shows the geometry, the energetics, and the Alfv\'enic Mach number of the jet. The reference field line, plotted in the upper left panel, shows that for most of the expansion the jet is parabolic with the height of the jet approximately the radius to the power of 4/3. At its highest point beyond the MFP, the jet has overcollimated to only 2 per cent of the maximum width, attained just before the MFP. The main difference between this solution and solution $c$ from PMM10 is that the height-to-width ratio has increased by a factor of 2.5, which means it has a narrower opening angle comparatively.

The upper right panel of Fig. \ref{overview} shows the partition of energy of the jet. After a small drop, the kinetic energy first increases at the expense of the thermal energy, signifying initial bulk acceleration due to a gas pressure gradient. After the flow has cooled, magnetic acceleration takes over and continues to accelerate the flow, also after overcollimation. Only just before the flow hits the axis does the Lorentz factor decrease again with a corresponding increase in the magnetic field strength.

The lower left panel of Fig. \ref{overview} shows the opening half-angle ($\pi/2 - \psi$) of the flow (also compare with Fig. \ref{alfvendistance}) and the causal connection opening angle. Both angles are measured from the axis of symmetry. The opening half-angle shows that after a relatively cylindrical start, the flow widens, before slowly collimating again. The causal connection opening angle is the equivalent of a sonic Mach cone. The flow cannot influence anything outside of the forward pointing cone with this angle. As the velocity increases, this cone becomes more narrow. This plot is almost identical to that for solution $c$.

The lower right panel of Fig. \ref{overview} shows the Alfv\'enic Mach number, the velocity in units of the Alfv\'en speed. After a long initial rise, shortly after overcollimation it decreases again. This decrease relates to the leftmost part of Fig. \ref{numden} where the numerator crosses zero again, without a corresponding crossing of the denominator, causing a decrease of the Mach number. Because the Mach number is defined in terms of the Alfv\'en speed, it is not the overall velocity that is decreasing, as there is no corresponding decrease in the Lorentz factor. It is rather the magnetic field strength that increases due to the overcollimation, leading to a higher Alfv\'en speed. The plot of the Mach number is very similar to that for solution $c$, but there is a twenty-fold increase of $M$. Note that in PMM10 we plotted the square of the Mach number.

\begin{figure*}
\includegraphics[width = 0.95 \textwidth]{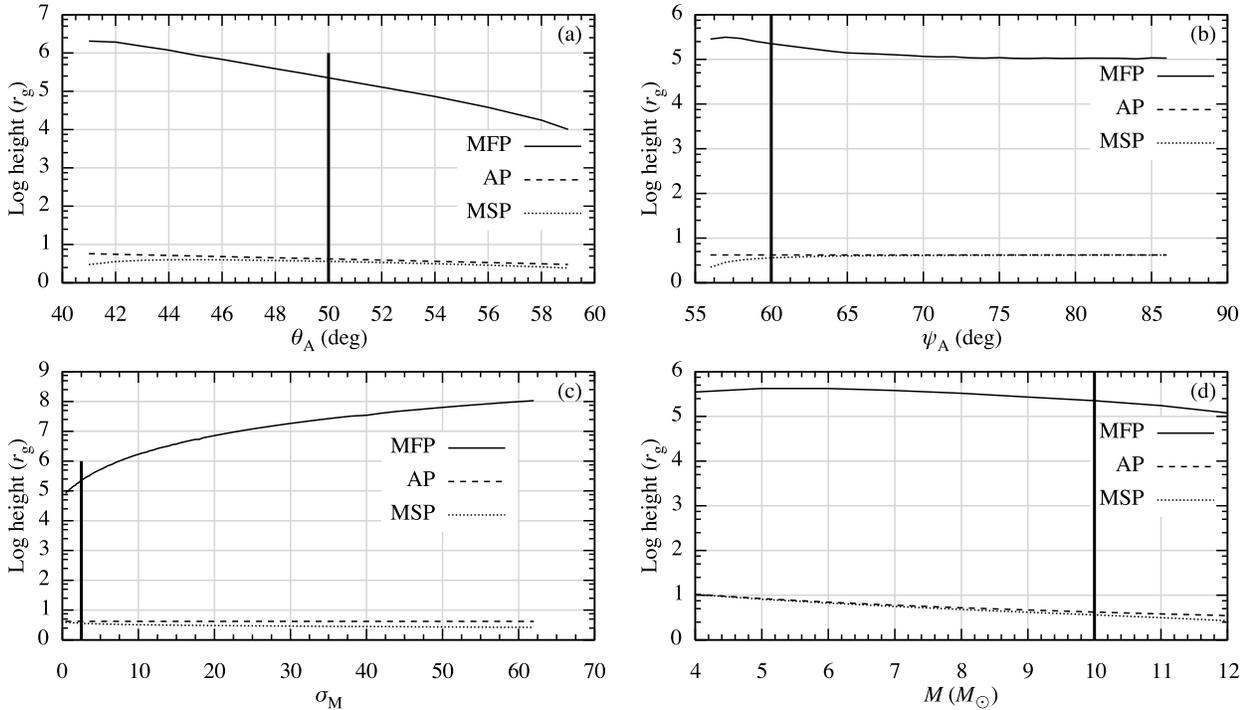}
\caption{
Examples of how the heights (in gravitational radii) of the solutions' three singular points change as each of the four principal free parameters is changed: solid/red line: MFP; dashed/blue line: AP; dotted/black line: MSP. The initial solution (see Table \ref{parametertable}) is indicated by the vertical black line. The parameters varied are as follows: panel (a): poloidal spherical angle of the Alfv\'en point ($\theta_\uA$); panel (b): the angle the field line makes with the disc at the Alfv\'en point ($\psi_\uA$); panel (c): magnetisation ($\sigma_\uM$); and panel (d): black hole mass ($\mathcal{M}$) in units of $M_\odot$. The horizontal axis in each of these plots is linear, not logarithmic. Note the dramatic excursions in the height of the MFP (3 orders of magnitude) with only modest changes in the free parameters. Note also that the MSP and AP heights are still outside the black hole horizon (see Fig. \ref{alfvendistance}); i.e., even though the height of the MSP $< 2~r_\ug$, the MSP spherical radius for these solutions remains outside the black hole horizon ($r > 2~r_\ug$). (A colour version of this figure is available in the online journal.)
}
\label{MFPMSPlocation}
\end{figure*}

\begin{figure*}
\includegraphics[width = 0.95 \textwidth]{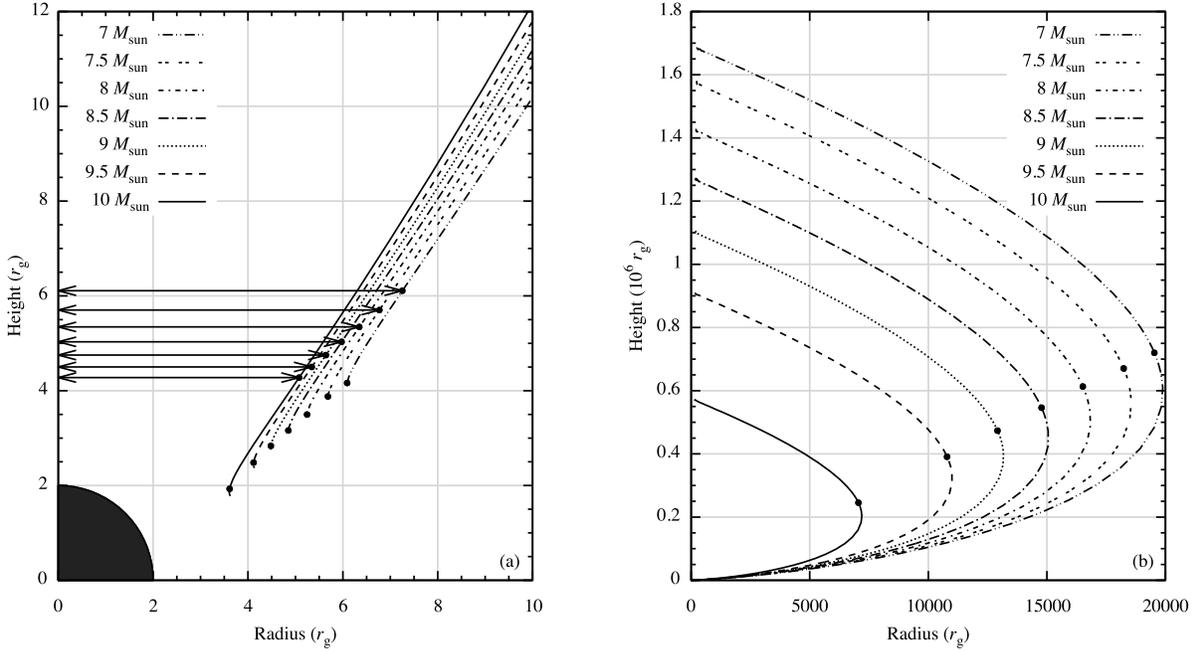}
\caption{
The effect of changing $\mathcal{M}$ on the field line geometry. The solid black/line is our canonical solution in Table \ref{parametertable}. $\mathcal{M}$ ranges from 7 -- 10~$M_\odot$ in steps of 0.5~$M_\odot$. In panel (a), the MSP and AP are indicated by the black dots. The arrows point to the AP. In panel (b) the MFP is indicated by black dots. Note that the height of the MSP (indicated at the lower end of the line) can be smaller than the Schwarzschild radius, but its spherical radius always remains outside the horizon. (A colour version of this figure is available in the online journal.)
}
\label{alfvendistance}
\end{figure*}

\subsection{An initial exploration of parameter space}
In this section we describe our exploration of parameter space. As a starting point we pick our first solution. We change the value of a single parameter ($\theta_\uA$, $\psi_\uA$, $\sigma_\uM$, or $\mathcal{M}$) until it is no longer possible to obtain a singular solution by fitting $x_\uA^2$, $q$, and $p_\uA$. The effect of changing these four parameters on the height of the MFP, AP and MSP is given in Fig. \ref{MFPMSPlocation}.

This plot also shows a physical reason why the solutions do not fill up the whole of parameter space. For high values of $\psi_\uA$, and for low values of $\sigma_\uM$ and $\mathcal{M}$, the MSP approaches the AP. Beyond the location where they coincide, no further solution is possible.

If $\theta_\uA$ and $\psi_\uA$ are changed together so their sum is roughly constant, the range of solutions spans approximately $45^\circ$. If they are changed separately, the range is significantly smaller. $\sigma_\uM$ has the largest spread for this case, ranging from 0.4 to above 60, monotonically increasing the height of the MFP. $\mathcal{M}$ has the range 4--12$~M_\odot$, spanning most of the mass range of astrophysical black holes for an AP distance of $\varpi_\uA \approx 74$ km.

Fig. \ref{MFPMSPlocation} gives the bounds for our solutions. Since we connect the start of the particle acceleration region with the MFP, we are mainly interested in its location within the jet. In the solutions found so far, the height of the MFP ranges from $10^4$ -- $10^8~r_\ug$. Because this work is by no means an exhaustive exploration of parameter space, this range provides only a preliminary indication. Nevertheless, it already allows us to model a variety of sources. The best way to change the height of the MFP seems to be adjusting $\theta_\uA$. The height monotonically decreases while increasing this parameter. Another possibility is decreasing $\sigma_\uM$. $\psi_\uA$ and $\mathcal{M}$ seem to have little effect over a wide range of values. The MSP should lie in or near the corona, thus locations too far away and too close to the black hole should be avoided. The height of the MSP ranges from 2 -- 10 $r_\ug$. This lower value is very close to the Schwarzschild radius and justifies the inclusion of the pseudo-Newtonian potential. The parameter $\mathcal{M}$ has the largest effect on the location of the MSP.

The effect of changing $\mathcal{M}$ on the geometry of the field lines is shown in Fig. \ref{alfvendistance}. This plot shows only the single reference field line of a solution. It is clear the height of the MSP can be smaller than the Schwarzschild radius, but the spherical radius can not. This geometry explains why the height of the MSP can be smaller than 2~$r_\ug$ in Fig. \ref{MFPMSPlocation}. As shown by the lower right panel of this figure, the solutions with a higher black hole mass collimate first, and reach a lower total height (in $r_\ug$).

\subsection{Self-similarity}
By including gravity in the relativistic equations of MHD, the assumption of self-similarity along the field line is broken. Near the black hole, the flow is non-relativistic and the wind equation reduces to the form of VTST00, which is self-similar. Far away from the centre, the gravitational energy is negligible and the wind equation assumes the form of VK03, which is also self-similar. This combination means that along the integrated field line, the flow transitions from one form of self-similarity to another. It should be noted that the reference field line we integrate will always be continuous.

Since one field line does not constitute a jet, we would like to be able to describe the flow in a region around our reference field line. Changing field lines is done by varying the parameter $\alpha = \varpi_\uA^2/\varpi_0^2$, where $\varpi_\uA$ is the cylindrical radius of the Alfv\'en point for a specific field line, and $\varpi_0$ is a reference length. By choosing this reference length to be the cylindrical radius of the Alfv\'en point for our reference field line, without loss of generality that field line has $\alpha = 1$.

Changing $\alpha$ to select different field lines then becomes equal to changing $\varpi_\uA$ or $\mathcal{M}$. Since this parameter is one of our model parameters, the other model parameters have to be fit in order to obtain a solutions passing through all three singular points. Although the best result would be obtained by changing all parameters simultaneously, for simplicity we limit ourselves to two. We are free to choose any two parameters, and when using $x_\uA^2$ and $q$ the field lines do not cross, which is required for self-similarity. Because $\theta_\uA$ and $\psi_\uA$ are therefore constant for all field lines in a region around our reference field line, at the AP self-similarity is exact. We will define this region, where the field lines do not cross and satisfy self-similarity to within a specified error, as a flux tube.

\begin{figure}
\includegraphics[angle = 270, width = 0.45 \textwidth]{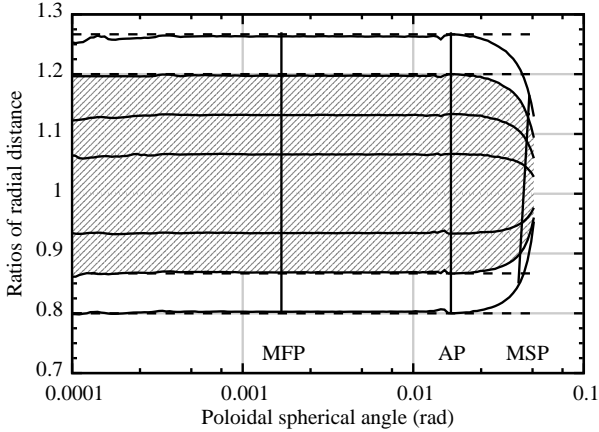}
\caption{
The ratios of the radial size of several field lines with regularly increasing $\varpi_\uA$ fitted with $x_\uA^2$ and $q$. The bottom line has $\alpha = 0.64$ and the top line has $\alpha = 1.6$ (where $\alpha \equiv \varpi_\uA^2/\varpi_0^2$). The horizontal dashed lines show the values for exact self-similarity. The deviations from self-similarity only get pronounced near the MSP. The shaded region shows the width of the flux tube for which the combined deviation is smaller than 10 per cent beyond the MSP. The parameters of the reference solution are $x_\uA^2 = 0.01$, $\sigma_\uM = 0.01$, $q = 0.01$, $\varpi_\uA = 5.08~r_\ug$, $\mathcal{M} = 10~M_\odot$, $\theta_\uA = 0.952915$, $\psi_\uA = 89.1119$. The Lorentz factor at the MFP is just above 11, showing this solution is relativistic. (A colour version of this figure is available in the online journal.)
}
\label{selfsimilarity}
\end{figure}

To show that the non-crossing of field lines holds for a large part of parameter space, we have performed this test at different parameter values. Fig. \ref{selfsimilarity} shows the deviations from self-similarity of this solution by dividing several field lines around a reference field line by this central field line. For perfect self-similarity this ratio should be a constant. In our case self-similarity is maintained for the majority of the jet. Only very close to the MSP do the deviations become more pronounced, due to gravity. These deviations are also demonstrated by the MSP occurring at slightly different angles for different values of the black hole mass $\mathcal{M}$ (see Fig. \ref{alfvendistance}). If we allow a maximum deviation of 10 per cent, which in this case will occur at the MSP since we will terminate our solutions there, the width of a flux tube would be about 0.30 of the radius of the central field line, which is a significant fraction.

Beyond the AP, deviations can be caused by changes in the fitting parameters to obtain singular solutions for the different field lines. The bigger the differences in these parameters, the greater the deviations and the narrower the resultant flux tube. The required changes in the parameters depend on the location in parameter space, and so the allowed width depends on the parameter values chosen. For the parameters of Fig. \ref{selfsimilarity}, the deviations beyond the AP are very small. While our equations do not strictly satisfy self-similarity, we have shown that it holds very well and that determining the width of the flux tube where it does is straightforward.

\section{Discussion}
\label{discussion}
By including gravity in the MHD equations, we are able to extend earlier solutions that crossed the MFP and AP to also go through the MSP, allowing us to describe a jet that crosses all three singular points. Because of this lower boundary condition, we now have a reliable description of the jet below the AP. This description gives a smooth solution from very near the central object out to the point of overcollimation, using a single partial differential equation to describe all regimes. This approach produces a workable physical model, allowing us to tie the jet properties to the conditions at the base, and providing a self-consistent determination of the start of the particle acceleration region.

At the same time, however, the addition of gravity also violates the conditions for self-similarity. The reason why a single self-similar relativistic flow equation, with gravity, has not been derived is because relativistic flow has one scaling with radius, while gravity has another. Our $C^\infty$-continuous bridging method has not changed that situation: while describable with a single, continuous equation, our solutions in the relativistic part of the flow without gravity will have different dependencies on the radius parameter $\alpha$ than in the non-relativistic part near the black hole that includes gravity. That is, at least one term in our wind equation will have a dependency on the radius parameter $\alpha$.

In effect our modified wind equation creates two regions with self-similar geometry, but with different self-similar dependencies in each region. Because our focus is on the bulk acceleration and collimation of jets in relativistic sources, our approach to this issue is to choose the self-similarity of the relativistic VK03 equations and restrict the different dependency on $\alpha$ to the gravity term only. That is, our solutions will not be strictly self-similar in the low-speed part of the flow with gravity (the VTST00 regime), but they will be self-similar far from the black hole in the VK03 regime (see Fig. \ref{selfsimilarity}). This highlights another advantage of this approach: since the gravity term is an algebraic one only (not involving any derivatives of the flow parameters with respect to either radius or polar angle), all the physical radial and angular dependence of the original equations will be preserved.

The different self-similarities within a single solution is not unique to our approach; it will be true of any method that attempts to bridge self-similar solutions with and without gravity. In this regard, interpretation of the solutions will be an important aspect of this study. There are two main issues to consider.

First, while accretion discs and jets may have a self-similar character, both simulations and observations show that their activity is concentrated in specific regions. For example, much of the accretion, and therefore jet, power is generated near the disc inner edge. And features like the one we seek (a strong collimation shock in the flow) will occur primarily at a specific point in the jet, and be observed as such, rather than being spread over a large range in radius. Therefore, we shall concentrate on only one bundle (`flux tube') of magnetic field and stream lines and assume that this flux tube is anchored near the disc inner edge and that it passes through the most important features of the jet. This will limit the range in $\alpha$ in which we are interested.

Second, in order that the solutions remain reasonably self-similar-like, we shall choose solutions in which the field lines do not cross within that flux tube. By choosing a specific combination of the free parameters to fit a solution with three singular points, the field lines around our reference field line behave well. We have found that selecting $x_\uA^2$ and $q$ as fitting parameters for the singular points produces results that satisfy self-similarity well within a flux tube of finite width.

With our method, therefore, we will restrict our solutions to a limited, and conservative, region of parameter space. The flow in our solutions must remain reasonably non-relativistic in the VTST00 part of the flow (when gravity is important), and we must consider only flux tubes of narrow enough width that field lines satisfy self-similarity to within a specified error in this region with gravity.

\begin{figure}
\includegraphics[width = 0.45 \textwidth]{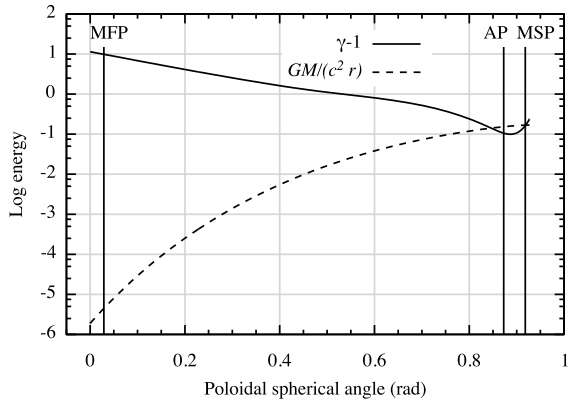}
\caption{
Comparison of the gravitational potential energy and the kinetic energy along the jet of the bottom solution detailed in Table \ref{parametertable}. With this plot we can select solutions where the gravitational energy $\mathcal{G M}/\left( c^2 r \right)$ is important close to the black hole, whereas the kinetic energy ($\gamma - 1$) only becomes relativistic further along the jet. (A colour version of this figure is available in the online journal.)
}
\label{energytest}
\end{figure}

Fig. \ref{energytest} can be used to see whether the assumptions of the two regimes are not violated for a particular solution. The kinetic energy ($\gamma - 1$) should be negligible near the black hole, while the gravitational potential energy $\left( \frac{\mathcal{G M}}{c^2 r} \right)$ should be significant, satisfying the conditions of the non-relativistic regime and demonstrating the importance of gravity. Around the Alfv\'en point the kinetic energy overtakes, without becoming relativistic. This is the bridging region where gravity becomes unimportant. Only beyond the Alfv\'en point does the flow become truly relativistic, while gravity becomes negligible, satisfying the conditions of the relativistic regime.

In this work we have only made a cursory exploration of the full parameter space, and it seems likely that part of that space will allow for solutions with smaller MFP values. The location of the MSP falls within the range of a radiatively-inefficient accretion flow (RIAF) disc model. Also, the velocity at the MSP, around $c/3$, is very reasonable. These features make it plausible that we will be able to match the conditions to an accretion flow as well as to the size of the jet base as indicated by fits to the spectrum \citep{2005ApJ...635.1203M}, allowing us to determine the location of the shock region from the conditions at the base of the jet.

It is important to note that since our assumption of non-relativistic flow (when gravity is important) eliminates the part of parameter space that produces relativistic flows close to the black hole, we exclude solutions that may relate to physical sources. Future work would include a more general model that allows relativistic flow in the gravitational field. However, the model in this paper provides a good description of the jet structure near the MSP when the flow is non-relativistic.

We should also note that the gravitational field we chose to include here is, at best, a pseudo-Newtonian (or pseudo-Schwarzschild) one. Our solutions, therefore, do not take into account any metric rotation that would occur near a Kerr black hole, for example. However, our method might be able to hint at the presence of a Kerr hole, if, for example, after fitting to broad-band data, we find that the footpoint of the field line may be significantly less than the ISCO of a Schwarzschild black hole ($\ll 6 \, r_\ug$) or that the required magnetisation is significantly greater than would be typical at the ISCO of a normal Keplerian disc.

Another issue neglected by our computations (and, indeed, by all steady state analyses) is the effect of instabilities on the accelerating jet flow.  We will briefly discuss fluid and MHD instabilities in the weak-field limit and then MHD instabilities in the strong-field limit as well. Figure 2b compares the relative strengths of fluid dynamical forces ({\it i.e.}, pressure $P$) with electromagnetic forces ({\it e.g.}, $(B^2 + E^2) / 8 \pi$).  We see that the former dominate only below the Alfv\'en point, in the slow magnetosonic region, so it is only there where we expect Kelvin-Helmholtz (KHI) or magneto-rotational instabilities (MRI) possibly to be important.  Indeed, the MSP itself may lie in the atmosphere of a turbulent accretion disk, giving rise to a somewhat more complex structure in the sub-Alfv\'enic region than that assumed here.  Nevertheless, recent two- and three-dimensional simulations of relativistic jets launched from turbulent accretion disks \citep{2006MNRAS.368.1561M,2009MNRAS.394L.126M} show that MHD jets similar to those investigated here (including a well-defined \emph{classical} slow point) do indeed form and are not disrupted by weak-field instabilities near the accretion disk.  Above this point, like other outflowing magnetospheres (solar, pulsar, {\it etc.}), the dynamical forces in the flow are so dominated by electromagnetic forces (eventually by many orders of magnitude) that any weak-field instabilities will be strongly suppressed, or at least unimportant in affecting the kinematics of the accelerating jet.

However, strong-field instabilities (in particular, the current-driven helical kink [CDI]) need to be looked at more closely. \citet{2004ApJ...617..123N}, for example, showed that in non-relativistic jet flow rotation velocities well above the Alfv\'en speed can suppress the helical kink, as can a steep external pressure gradient.  They speculated that relativistic flow would further reduce the development of kinks, but a similar detailed relativistic study has not yet been performed.  Again, as one of the few relativistic three-dimensional MHD simulations, \citet{2009MNRAS.394L.126M} can give some insight into the behavior of real MHD jets in the strong-field region.  While the accelerating jet appears to be affected somewhat by helical kinking in that study, it does not appear to be disrupted by the CDI.  It remains a viable jet well beyond the classical fast point (where the rotational speed should be significantly super-Alfv\'enic.  However, no three-dimensional simulation so far has followed the acceleration out to the MFP region, let alone for long model times to achieve a quasi-steady state. And, detailed numerical parameter studies are even further in the future.  So, the question of whether MHD jet acceleration can be largely stable to strong-field MHD instabilities is still an open question.

\section{Conclusions}
We have shown that it is possible to extend the time independent, semi-analytic solutions of PMM10 to solutions that include gravity using a $C^\infty$-continuous bridging method, connecting two valid regimes with a smooth transition region. We also extend the MHD wind/jet equation to include a pseudo-Newtonian potential for gravity. For the first time these solutions cross all three singular points and describe a relativistic jet from the vicinity of a central black hole to the point where the flow hits the axis of symmetry. Although these solutions do not satisfy strict self-similarity, the deviations within a flux tube of a certain width can be estimated and controlled, and thus a physically relevant model can be constructed.

We have discovered a solution crossing all three singular points with parameters very close to one from our previous paper. This new solution shows the relative ease with which solutions can be found and allows for a comparison. The main differences are the higher distance of the MFP and a significantly higher velocity, both owing to a higher bulk acceleration at the AP. Otherwise the two solutions look very similar, showing that gravity has little influence beyond the region close to the central object. By allowing the creation of an MSP, including gravity enables us to have a reliable description of the flow below the AP.

We have made a cursory examination of parameter space by changing one parameter of a single solution at a time. This exploration gives a preliminary indication of the extent of the solution space, while at the same time showing the effects it has on the solutions. In particular we have found that the poloidal spherical angle of the AP has the most significant effect on the height of the MFP. On the other hand, the ratio of the AP to the light cylinder radius has the greatest effect on the height of the MSP.

The multi-dimensional solution space is constrained by physical and mathematical conditions. Changing only one parameter at a time, when the light cylinder radius approaches the AP, the AP moves outwards, the temperature is increased, or the magnetisation decreased, the jets become wider and shorter, with the MSP approaching the AP. By increasing the cylindrical radius of the AP, the effects of gravity decrease. Eventually this decrease leads to the regime where an MSP cannot be created anymore, approaching the situation without gravity. Despite all these constraints, the solution space allows a wide range of parameter values to be chosen, translating into a wide range of properties, like the location of the MSP and MFP, the velocity, magnetic energy, density and pressure of the flow.

By matching conditions at the MSP to an accretion flow model, the mathematical parameters considered free in this work will be tied to the conditions at the base of the jet and subsumed into the model, with the wide range of properties in the solution space assuring a good fit. Consequently, the height of the MFP, and the start of the particle acceleration region, will be uniquely determined by the conditions close to the central object, providing a self-consistent connection. After integration into a model that can determine the spectral emission of a jet solution, we will be able to ascertain the conditions that best describe the overall spectrum of a given black hole system. In future work we will explore how well this model succeeds in predicting the correct location of the optically thick-to-thin break observed in the broadband spectra of compact jets, hopefully with new insights about the physics of jet launching conditions.

\section*{Acknowledgments}
P. Polko and S. Markoff gratefully acknowledge support from a Netherlands Organisation for Scientific Research (NWO) Vidi Fellowship. In addition, S. Markoff is grateful for support from the European Community's Seventh Framework Program (FP7/2007-2013) under grant agreement number ITN 215212 `Black Hole Universe'. Part of the research described in this paper was carried out at the Jet Propulsion Laboratory, California Institute of Technology, under a contract with the National Aeronautics and Space Administration.

\appendix
\section{Derivation of the gravity term}
\label{appendix:gravity}
To obtain a relativistic form of gravity, we need the following conversions: $\varpi_*$ in VTST00 is equal to $\varpi_0$ in VK03 and $V_*$ similarly corresponds to $\frac{\alpha^{1/4} F \sigma_\uM c}{\gamma \xi x_\uA^2}$.

Substituting these into equation (17) of VTST00 produces:
\begin{align}
\kappa &= \sqrt{\frac{\mathcal{G M}}{\varpi_* V_*^2}} = \sqrt{\frac{\mathcal{G M} \gamma^2 \xi^2 x_\uA^4}{\varpi_0 \alpha^{1/2} F^2 \sigma_\uM^2 c^2}} \nonumber \\
&= \sqrt{\frac{\mathcal{G M}}{c^2 \varpi_\uA} \frac{\mu^2 x_\uA^4}{F^2 \sigma_\uM^2} \frac{(1 - M^2 - x_\uA^2)^2}{(1 - M^2 - x^2)^2}}.
\label{kappa}
\end{align}
The full gravity term then becomes:
\begin{equation}
- \frac{\kappa^2 \sin(\theta)}{G} = - \frac{\mathcal{G M}}{c^2} \frac{\mu^2 x_\uA^4}{F^2 \sigma_\uM^2} \frac{(1 - M^2 - x_\uA^2)^2}{(1 - M^2 - x^2)^2} \frac{\sin(\theta)}{\varpi_\uA G}.
\end{equation}
In this expression $(\varpi_\uA G)/\sin(\theta)$ is equal to the spherical radius, $r$. To include a pseudo-Newtonian potential, we replace $1/r$ by $1/(r - r_\uS)$, where $r_\uS$ is the Schwarzschild radius ($= 2 \mathcal{G M}/c^2$) and simplify:
\begin{equation}
- \frac{\kappa^2 \sin(\theta)}{G} = - \frac{\mu^2 x_\uA^4}{F^2 \sigma_\uM^2} \frac{(1 - M^2 - x_\uA^2)^2}{(1 - M^2 - x^2)^2} \left[ \frac{c^2 \varpi_\uA G}{\mathcal{G M} \sin(\theta)} - 2 \right]^{-1}.
\end{equation}

\section{Equations for the initial parameter values}
The Alfv\'en regularity condition is given by evaluating the wind equation at the AP:
\begin{align}
& \frac{F^2 \sigma_\uM^2 (1 - x_\uA^2) (\sigma_\uA + 1)^2 \sin(\theta_\uA)}{\mu^2 \cos^2(\theta_\uA + \psi_\uA)} \bigg\{ \nonumber \\
& - 2 \frac{\Gamma - 1}{\Gamma} \frac{(F - 2) ( \xi_\uA - 1) (1 - x_\uA^2)}{\xi_\uA x_\uA^2} \sin(\theta_\uA) \nonumber \\
& + 2 \cos(\psi_\uA) \sin(\theta_\uA + \psi_\uA) \frac{\sigma_\uA + 1}{\sigma_\uA} \nonumber \\
& + \frac{\sin(\theta_\uA)}{x_\uA^2} \left[ (F - 1) (1 - x_\uA^2) - 1\right] \bigg\} \nonumber \\
= {} & \left[ x_\uA^2 - \sigma_\uA (1 - x_\uA^2) \right]^2 - (F - 1) \sigma_\uA^2 (1 - x_\uA^2) \nonumber \\
& - 2 \frac{\Gamma - 1}{\Gamma} (F - 2) \frac{\xi_\uA - 1}{\xi_\uA} \left\{ x_\uA^2 - \left[ x_\uA^2 - \sigma_\uA (1 - x_\uA^2) \right]^2 \right\} \nonumber \\
& - \frac{x_\uA^2}{1 - x_\uA^2} \left[ \frac{c^2 \varpi_\uA}{\mathcal{G M} \sin(\theta_\uA)} - 2 \right]^{-1}
\label{ARC}
\end{align}
We obtain $\mu^2$ by evaluating the Bernoulli equation at the AP:
\begin{align}
\mu^2 = {} & \frac{(\sigma_\uA+1)^2}{x_\uA^2 - \left[ x_\uA^2 - \sigma_\uA \left( 1 - x_\uA^2 \right) \right]^2} \bigg[ x_\uA^2 \xi_\uA^2 \nonumber \\
& + \frac{F^2 \sigma_\uM^2 \left( 1 - x_\uA^2 \right)^2 \sin^2(\theta_\uA)}{x_\uA^2 \cos^2(\theta_\uA + \psi_\uA)} \bigg]
\label{mu2}
\end{align}

\section{Gravity in the Bernoulli equation}
\label{Bernoulli}
The only terms that include the effects of gravity in the Bernoulli equation given by (A3) in VTST00 are the Bernoulli constant and the gravity term. Dividing out a common factor of 2, we have:
\begin{equation}
\varepsilon + \frac{\kappa^2 \sin(\theta)}{G}.
\end{equation}
At the AP $G = 1$ and $\theta = \theta_\uA$, reducing these terms to:
\begin{equation}
\varepsilon + \kappa^2 \sin(\theta_\uA).
\label{gravityterms}
\end{equation}
If we write out $\varepsilon$ using equation (19) of VTST00, $G \equiv \varpi / \varpi_\uA$, and equation (2.7a) of BP82, we obtain (please note that to keep to the notation of the preceding two papers, here the subscript 0 means the value at the base of the outflow, not the reference values as used in VK03. Unfortunately, $r_0$ as used in BP82, is $\varpi_0$ as used in VTST00):
\begin{equation}
\varepsilon = \varepsilon \frac{\kappa^2}{G_0} = \frac{e}{\mathcal{G M} / \varpi_0} \frac{\kappa^2}{\varpi_0 / \varpi_\uA} = \kappa^2 \frac{e}{\mathcal{G M} / \varpi_\uA}.
\label{Bernoulliconstant}
\end{equation}
Since $e$, the specific energy, is a constant of motion, we can evaluate it at the AP, just like the Bernoulli equation in VK03 (equation (2.2) of BP82 with the gravitational potential expanded):
\begin{equation}
e = e_\uA = \frac{V_\uA^2}{2} + h_\uA - \frac{\mathcal{G M}}{r_\uA} - \frac{\Omega \varpi_\uA B_{\phi,\uA}}{\Psi_{A,\uA}}.
\label{specificenergy}
\end{equation}
Here $r_\uA$ is the spherical radius of the AP. Substituting \eqref{specificenergy} and \eqref{Bernoulliconstant} into \eqref{gravityterms}, we obtain:
\begin{equation}
\kappa^2 \left[ \frac{\varpi_\uA}{\mathcal{G M}} \left( \frac{V_\uA^2}{2} + h_\uA - \frac{\Omega \varpi_\uA B_{\phi,\uA}}{\Psi_{A,\uA}} \right) - \frac{\varpi_\uA}{r_\uA} + \sin(\theta_\uA) \right].
\end{equation}
However, $\varpi_\uA / r_\uA = \sin(\theta)$, so the last two terms cancel. The $\kappa^2$ term cancels the $\mathcal{G M}$ factor (equation \ref{kappa}), leaving no terms that depend on gravity. Thus there is no dependence on gravity in the Bernoulli equation in the non-relativistic regime with gravity at the AP and consequently there is no gravity term in the relativistic Bernoulli equation at the AP (equation \ref{mu2}).

\label{lastpage}


\begin{thebibliography}{}

\bibitem[\protect\citeauthoryear{{Asada}, {Nakamura}, {Doi}, {Nagai} \&
  {Inoue}}{{Asada} et~al.}{2011}]{2011IAUS..275..198A}
{Asada} K.,  {Nakamura} M.,  {Doi} A.,  {Nagai} H.,    {Inoue} M.,  2011, in
  {G.~E.~Romero, R.~A.~Sunyaev, \& T.~Belloni} ed., IAU Symposium Vol.~275 of
  IAU Symposium, {EVN monitoring observation of M 87 jet}.
pp 198--199

\bibitem[\protect\citeauthoryear{{Bell}}{{Bell}}{1978}]{1978MNRAS.182..147B}
{Bell} A.~R.,  1978, \mnras, 182, 147

\bibitem[\protect\citeauthoryear{{Blandford} \& {Begelman}}{{Blandford} \&
  {Begelman}}{1999}]{1999MNRAS.303L...1B}
{Blandford} R.~D.,  {Begelman} M.~C.,  1999, \mnras, 303, L1

\bibitem[\protect\citeauthoryear{{Blandford} \& {Konigl}}{{Blandford} \&
  {Konigl}}{1979}]{1979ApJ...232...34B}
{Blandford} R.~D.,  {Konigl} A.,  1979, \apj, 232, 34

\bibitem[\protect\citeauthoryear{{Blandford} \& {Payne}}{{Blandford} \&
  {Payne}}{1982}]{1982MNRAS.199..883B}
{Blandford} R.~D.,  {Payne} D.~G.,  1982, \mnras, 199, 883

\bibitem[\protect\citeauthoryear{{Corbel} \& {Fender}}{{Corbel} \&
  {Fender}}{2002}]{2002ApJ...573L..35C}
{Corbel} S.,  {Fender} R.~P.,  2002, \apjl, 573, L35

\bibitem[\protect\citeauthoryear{{Drury}}{{Drury}}{1983}]{1983RPPh...46..973D}
{Drury} L.~O.,  1983, Reports on Progress in Physics, 46, 973

\bibitem[\protect\citeauthoryear{{Esin}, {McClintock} \& {Narayan}}{{Esin}
  et~al.}{1997}]{1997ApJ...489..865E}
{Esin} A.~A.,  {McClintock} J.~E.,    {Narayan} R.,  1997, \apj, 489, 865

\bibitem[\protect\citeauthoryear{{Falcke} \& {Biermann}}{{Falcke} \&
  {Biermann}}{1995}]{1995A&A...293..665F}
{Falcke} H.,  {Biermann} P.~L.,  1995, \aap, 293, 665

\bibitem[\protect\citeauthoryear{{Falcke}, {K{\"o}rding} \& {Markoff}}{{Falcke}
  et~al.}{2004}]{2004A&A...414..895F}
{Falcke} H.,  {K{\"o}rding} E.,    {Markoff} S.,  2004, \aap, 414, 895

\bibitem[\protect\citeauthoryear{{Falcke} \& {Markoff}}{{Falcke} \&
  {Markoff}}{2000}]{2000A&A...362..113F}
{Falcke} H.,  {Markoff} S.,  2000, \aap, 362, 113

\bibitem[\protect\citeauthoryear{{Fender}}{{Fender}}{2001}]{2001MNRAS.322...31F}
{Fender} R.~P.,  2001, \mnras, 322, 31

\bibitem[\protect\citeauthoryear{{Fender}, {Belloni} \& {Gallo}}{{Fender}
  et~al.}{2004}]{2004MNRAS.355.1105F}
{Fender} R.~P.,  {Belloni} T.~M.,    {Gallo} E.,  2004, \mnras, 355, 1105

\bibitem[\protect\citeauthoryear{{Gallo}, {Migliari}, {Markoff}, {Tomsick},
  {Bailyn}, {Berta}, {Fender} \& {Miller-Jones}}{{Gallo}
  et~al.}{2007}]{2007ApJ...670..600G}
{Gallo} E.,  {Migliari} S.,  {Markoff} S.,  {Tomsick} J.~A.,  {Bailyn} C.~D.,
  {Berta} S.,  {Fender} R.,    {Miller-Jones} J.~C.~A.,  2007, \apj, 670, 600

\bibitem[\protect\citeauthoryear{{Gandhi}, {Blain}, {Russell}, {Casella},
  {Malzac}, {Corbel}, {D'Avanzo}, {Lewis}, {Markoff}, {Cadolle Bel}, {Goldoni},
  {Wachter}, {Khangulyan} \& {Mainzer}}{{Gandhi}
  et~al.}{2011}]{2011ApJ...740L..13G}
{Gandhi} P.,  {Blain} A.~W.,  {Russell} D.~M.,  {Casella} P.,  {Malzac} J.,
  {Corbel} S.,  {D'Avanzo} P.,  {Lewis} F.~W.,  {Markoff} S.,  {Cadolle Bel}
  M.,  {Goldoni} P.,  {Wachter} S.,  {Khangulyan} D.,    {Mainzer} A.,  2011,
  \apjl, 740, L13

\bibitem[\protect\citeauthoryear{{Gebhardt} \& {Thomas}}{{Gebhardt} \&
  {Thomas}}{2009}]{2009ApJ...700.1690G}
{Gebhardt} K.,  {Thomas} J.,  2009, \apj, 700, 1690

\bibitem[\protect\citeauthoryear{{Hada}, {Doi}, {Kino}, {Nagai}, {Hagiwara} \&
  {Kawaguchi}}{{Hada} et~al.}{2011}]{2011Natur.477..185H}
{Hada} K.,  {Doi} A.,  {Kino} M.,  {Nagai} H.,  {Hagiwara} Y.,    {Kawaguchi}
  N.,  2011, \nat, 477, 185

\bibitem[\protect\citeauthoryear{{Heinz} \& {Sunyaev}}{{Heinz} \&
  {Sunyaev}}{2003}]{2003MNRAS.343L..59H}
{Heinz} S.,  {Sunyaev} R.~A.,  2003, \mnras, 343, L59

\bibitem[\protect\citeauthoryear{{Ho}}{{Ho}}{1999}]{1999ApJ...516..672H}
{Ho} L.~C.,  1999, \apj, 516, 672

\bibitem[\protect\citeauthoryear{{Junor}, {Biretta} \& {Livio}}{{Junor}
  et~al.}{1999}]{1999Natur.401..891J}
{Junor} W.,  {Biretta} J.~A.,    {Livio} M.,  1999, \nat, 401, 891

\bibitem[\protect\citeauthoryear{{Li}, {Chiueh} \& {Begelman}}{{Li}
  et~al.}{1992}]{1992ApJ...394..459L}
{Li} Z.-Y.,  {Chiueh} T.,    {Begelman} M.~C.,  1992, \apj, 394, 459

\bibitem[\protect\citeauthoryear{{Maitra}, {Markoff}, {Brocksopp}, {Noble},
  {Nowak} \& {Wilms}}{{Maitra} et~al.}{2009}]{2009MNRAS.398.1638M}
{Maitra} D.,  {Markoff} S.,  {Brocksopp} C.,  {Noble} M.,  {Nowak} M.,
  {Wilms} J.,  2009, \mnras, 398, 1638

\bibitem[\protect\citeauthoryear{{Markoff}, {Falcke} \& {Fender}}{{Markoff}
  et~al.}{2001}]{2001A&A...372L..25M}
{Markoff} S.,  {Falcke} H.,    {Fender} R.,  2001, \aap, 372, L25

\bibitem[\protect\citeauthoryear{{Markoff}, {Nowak}, {Corbel}, {Fender} \&
  {Falcke}}{{Markoff} et~al.}{2003}]{2003A&A...397..645M}
{Markoff} S.,  {Nowak} M.,  {Corbel} S.,  {Fender} R.,    {Falcke} H.,  2003,
  \aap, 397, 645

\bibitem[\protect\citeauthoryear{{Markoff}, {Nowak}, {Young}, {Marshall},
  {Canizares}, {Peck}, {Krips}, {Petitpas}, {Sch{\"o}del}, {Bower}, {Chandra},
  {Ray}, {Muno}, {Gallagher}, {Hornstein} \& {Cheung}}{{Markoff}
  et~al.}{2008}]{2008ApJ...681..905M}
{Markoff} S.,  {Nowak} M.,  {Young} A.,  {Marshall} H.~L.,  {Canizares} C.~R.,
  {Peck} A.,  {Krips} M.,  {Petitpas} G.,  {Sch{\"o}del} R.,  {Bower} G.~C.,
  {Chandra} P.,  {Ray} A.,  {Muno} M.,  {Gallagher} S.,  {Hornstein} S.,
  {Cheung} C.~C.,  2008, \apj, 681, 905

\bibitem[\protect\citeauthoryear{{Markoff}, {Nowak} \& {Wilms}}{{Markoff}
  et~al.}{2005}]{2005ApJ...635.1203M}
{Markoff} S.,  {Nowak} M.~A.,    {Wilms} J.,  2005, \apj, 635, 1203

\bibitem[\protect\citeauthoryear{{Marscher} \& {Gear}}{{Marscher} \&
  {Gear}}{1985}]{1985ApJ...298..114M}
{Marscher} A.~P.,  {Gear} W.~K.,  1985, \apj, 298, 114

\bibitem[\protect\citeauthoryear{{McClintock} \& {Remillard}}{{McClintock} \&
  {Remillard}}{2006}]{2006csxs.book..157M}
{McClintock} J.~E.,  {Remillard} R.~A.,  2006, {Black hole binaries}.
Cambridge University Press

\bibitem[\protect\citeauthoryear{{McKinney}}{{McKinney}}{2006}]{2006MNRAS.368.1561M}
{McKinney} J.~C.,  2006, \mnras, 368, 1561

\bibitem[\protect\citeauthoryear{{McKinney} \& {Blandford}}{{McKinney} \&
  {Blandford}}{2009}]{2009MNRAS.394L.126M}
{McKinney} J.~C.,  {Blandford} R.~D.,  2009, \mnras, 394, L126

\bibitem[\protect\citeauthoryear{{Meier}}{{Meier}}{2001}]{2001ApJ...548L...9M}
{Meier} D.~L.,  2001, \apjl, 548, L9

\bibitem[\protect\citeauthoryear{{Merloni} \& {Fabian}}{{Merloni} \&
  {Fabian}}{2002}]{2002MNRAS.332..165M}
{Merloni} A.,  {Fabian} A.~C.,  2002, \mnras, 332, 165

\bibitem[\protect\citeauthoryear{{Merloni}, {Heinz} \& {di Matteo}}{{Merloni}
  et~al.}{2003}]{2003MNRAS.345.1057M}
{Merloni} A.,  {Heinz} S.,    {di Matteo} T.,  2003, \mnras, 345, 1057

\bibitem[\protect\citeauthoryear{{Migliari}, {Tomsick}, {Maccarone}, {Gallo},
  {Fender}, {Nelemans} \& {Russell}}{{Migliari}
  et~al.}{2006}]{2006ApJ...643L..41M}
{Migliari} S.,  {Tomsick} J.~A.,  {Maccarone} T.~J.,  {Gallo} E.,  {Fender}
  R.~P.,  {Nelemans} G.,    {Russell} D.~M.,  2006, \apjl, 643, L41

\bibitem[\protect\citeauthoryear{{Migliari}, {Tomsick}, {Markoff}, {Kalemci},
  {Bailyn}, {Buxton}, {Corbel}, {Fender} \& {Kaaret}}{{Migliari}
  et~al.}{2007}]{2007ApJ...670..610M}
{Migliari} S.,  {Tomsick} J.~A.,  {Markoff} S.,  {Kalemci} E.,  {Bailyn} C.~D.,
   {Buxton} M.,  {Corbel} S.,  {Fender} R.~P.,    {Kaaret} P.,  2007, \apj,
  670, 610

\bibitem[\protect\citeauthoryear{{Nakamura} \& {Meier}}{{Nakamura} \&
  {Meier}}{2004}]{2004ApJ...617..123N}
{Nakamura} M.,  {Meier} D.~L.,  2004, \apj, 617, 123

\bibitem[\protect\citeauthoryear{{Narayan} \& {Yi}}{{Narayan} \&
  {Yi}}{1994}]{1994ApJ...428L..13N}
{Narayan} R.,  {Yi} I.,  1994, \apjl, 428, L13

\bibitem[\protect\citeauthoryear{{Paczy{\'n}sky} \& {Wiita}}{{Paczy{\'n}sky} \&
  {Wiita}}{1980}]{1980A&A....88...23P}
{Paczy{\'n}sky} B.,  {Wiita} P.~J.,  1980, \aap, 88, 23

\bibitem[\protect\citeauthoryear{{Plotkin}, {Markoff}, {Kelly}, {K{\"o}rding}
  \& {Anderson}}{{Plotkin} et~al.}{2012}]{2012MNRAS.419..267P}
{Plotkin} R.~M.,  {Markoff} S.,  {Kelly} B.~C.,  {K{\"o}rding} E.,
  {Anderson} S.~F.,  2012, \mnras, 419, 267

\bibitem[\protect\citeauthoryear{{Polko}, {Meier} \& {Markoff}}{{Polko}
  et~al.}{2010}]{2010ApJ...723.1343P}
{Polko} P.,  {Meier} D.~L.,    {Markoff} S.,  2010, \apj, 723, 1343 (Paper I)

\bibitem[\protect\citeauthoryear{{Quataert} \& {Gruzinov}}{{Quataert} \&
  {Gruzinov}}{2000}]{2000ApJ...539..809Q}
{Quataert} E.,  {Gruzinov} A.,  2000, \apj, 539, 809

\bibitem[\protect\citeauthoryear{{Rahoui}, {Lee}, {Heinz}, {Hines},
  {Pottschmidt}, {Wilms} \& {Grinberg}}{{Rahoui}
  et~al.}{2011}]{2011ApJ...736...63R}
{Rahoui} F.,  {Lee} J.~C.,  {Heinz} S.,  {Hines} D.~C.,  {Pottschmidt} K.,
  {Wilms} J.,    {Grinberg} V.,  2011, \apj, 736, 63

\bibitem[\protect\citeauthoryear{{Shakura} \& {Sunyaev}}{{Shakura} \&
  {Sunyaev}}{1973}]{1973A&A....24..337S}
{Shakura} N.~I.,  {Sunyaev} R.~A.,  1973, \aap, 24, 337

\bibitem[\protect\citeauthoryear{{Tonry}, {Dressler}, {Blakeslee}, {Ajhar},
  {Fletcher}, {Luppino}, {Metzger} \& {Moore}}{{Tonry}
  et~al.}{2001}]{2001ApJ...546..681T}
{Tonry} J.~L.,  {Dressler} A.,  {Blakeslee} J.~P.,  {Ajhar} E.~A.,  {Fletcher}
  A.~B.,  {Luppino} G.~A.,  {Metzger} M.~R.,    {Moore} C.~B.,  2001, \apj,
  546, 681

\bibitem[\protect\citeauthoryear{{Vlahakis} \& {K{\"o}nigl}}{{Vlahakis} \&
  {K{\"o}nigl}}{2003}]{2003ApJ...596.1080V}
{Vlahakis} N.,  {K{\"o}nigl} A.,  2003, \apj, 596, 1080

\bibitem[\protect\citeauthoryear{{Vlahakis}, {Tsinganos}, {Sauty} \&
  {Trussoni}}{{Vlahakis} et~al.}{2000}]{2000MNRAS.318..417V}
{Vlahakis} N.,  {Tsinganos} K.,  {Sauty} C.,    {Trussoni} E.,  2000, \mnras,
  318, 417

\bibitem[\protect\citeauthoryear{{Walker}, {Ly}, {Junor} \& {Hardee}}{{Walker}
  et~al.}{2008}]{2008JPhCS.131a2053W}
{Walker} R.~C.,  {Ly} C.,  {Junor} W.,    {Hardee} P.~J.,  2008, Journal of
  Physics Conference Series, 131, 012053

\end{thebibliography}
\end{document}